\documentclass[12pt]{iopart}

\usepackage{graphicx}
\begin{document}

\title[Quantum discord and 1-norm GQD for XXX Heisenberg chain under noisy channels ]{Quantum discord and one-norm geometric quantum discord for XXX Heisenberg spin chain under noisy channels}

\author{E. Faizi, H. Eftekhari}

\address{Physics Department, Azarbaijan shahid madani university}
\ead{ efaizi@azaruniv.edu}
\begin{abstract}
 Here, we compare the behavior of the quantum discord, 1- norm geometric quantum discord and concurrence under the effect of decoherence for XXX Heisenberg spin chain. Their dependencies on temperature and coupling constant are presented in detail. By a comparison between them we can see that the quantum discord and 1 -norm geometric quantum discord are more robust than the concurrence against decoherence in the sense that quantum discord and 1- norm geometric quantum discord takes a nonzero value and decrease asymptotically, while the concurrence is zero in a large range of the parameters including temperature and coupling constant J. Moreover, we find that  there is no simple relative ordering between 1- norm geometric quantum discord and quantum discord. This result contradicts the information which provided in some papers (see ref. \cite{Paula}).

\end{abstract}


\section{Introduction}
Quantum discord (QD) has attracted much attention, since it was proposed by Ollivier and Zurek \cite{Ollivier, K. Modi}. It specifies the quantumness of correlations in quantum states from a measurement aspect and is basically different from quantum entanglement. It has been demonstrated theoretically \cite{Animesh} and experimentally \cite{Lanyon}, in deterministic quantum computation with one pure qubit (DQC1), nonzero QD without unentangled states is responsible for computation speedup. QD has essential role in quantum information tasks, thus it is unavoidable to study the properties of QD. The geometrization of QD, namely geometric quantum discord (GQD), has been introduced by Dakic, Vedral, and Brukner \cite{Brukner}, mainly stimulated by the difficulty of finding analytical solutions for the entropic version of QD. GQD, quantifies the amount of quantum correlations in terms of its minimal Hilbert- Schmidt distance from the set of classical states \cite{Costa}. It is computable for general two- qubit states analytically \cite{Brukner} also for any arbitrary bipartite states\cite{Luo,Rana}. In addition, it has operational significance in certain quantum communication tasks (see, e.g.,Refs. \cite{Dakic,Gu}). In spite of those remarkable features, GQD is sensitive to the choice of distance measures (see, e.g., ref. \cite{Bellomo1}). In turn, as recently indicated \cite{Hu, Tufarelli}, GQD as proposed in ref. \cite{Brukner} cannot be a good measure for the quantumness of correlations, because it may increase under local operations on the unmeasured subsystem. Specifically, it has been pointed out by Piani  that the introduction of a factorized local ancillary state on the unmeasured party changes the geometric quantum discord by a factor given by purity of the ancilla \cite{Piani}. However, this problem can be worked out if Schatten 1- norm (trace norm) is utilized as a distance measure \cite{B. Aaronson, Aaronson, T. Nakano}.
\\In recent years, with the progress of quantum information and quantum computation, spin system applies in the field of quantum information widely \cite{Ma, Guo}. Heisenberg model (as the simplest spin chain) has been investigated in many subjects of quantum information and computation. It can be applied in many
physical systems such as quantum dot system \cite{Kane}, nucleus system \cite{Vrijen}, electronic spin system \cite{Sorensen} and optical lattices system \cite{Lidar} and so on. Quantum correlation in the Heisenberg
model make a connection between quantum information and condensed matter physics.\\
 In the dynamics of any actual open quantum system we have encountered with the various kind of interactions between system and its environment.  These interaction which can lead to decoherence and change the quantum properties \cite{H. P. Breuer, T. Yu, R. Lo}. These interactions typically lead to decoherence and change the quantum properties. Entanglement and QD as two distinct features of quantum correlations have fundamental roles in executing quantum information tasks, so that the investigation of their decoherence dynamics is important \cite{K. Modi, R, R. Horodecki, B, R. Lo Franco,  B. Bellomo, Bellomo, M. L. Hu}. For example, Werlang et al. \cite{Werlang} proposed that QD is more robust than entanglement against decoherence in Markovian environments. QD has been shown to be more resilient than entanglement also in non- Markovian quantum and classical environments, both theoretically \cite{X, Y} and experimentally \cite{J. S. Xu}.

In the present paper, we deal with 1D Heisenberg open chains with no external magnetic field and only nearest neighbor interactions. Here properties of QD, 1-norm GQD and entanglement in the Heisenberg XXX spin chain are investigated. How these different aspects of quantum correlation varies in the presence of bit flip (BF) and generalized amplitude damping (GAD) noises environment is also revealed. In the studies of the effect of noises on quantum discord, some studies \cite{Werlang} have generally been allocated to explore only the time evolution of quantum discord.  But, here we study the dependence of quantum discord and the 1- norm GQD to temperature  for XXX model. In addition, we analyze the QD and 1- norm GQD in both ferromagnetic and anti- ferromagnetic cases. The rest of this paper is organized as follows. In sect. 2 and 3 we briefly review the definition of QD and schatten 1-norm GQD. Also, in sect. 4 We  present the model of Heisenberg XXX chain with  no external magnetic
field and only nearest neighbor interactions. In sect. 5 and 6 we discuss the quantum correlation without decoherence and with decoherence respectively. Conclusion is given in sect. 7.

\section{Quantum discord and entanglement}
In classical information theory, the total correlation between two parts can be represented by two kinds of equivalent expressions of mutual information. In the quantum territory, one of quantum expansion of mutual information is equal to total correlation. It can be written as:
\begin{eqnarray} I({\rho}_ {AB})=S({\rho}_A)+S({\rho}_B)-S({\rho}_{AB}),
\end{eqnarray}
where $S({\rho})=-Tr(\rho\log_2\rho)$ is Von Neumann entropy and  ${\rho}_A({\rho}_B)$ is the reduced density operator of the section A(B). Another quantum expression for mutual information can be obtained after a complete set of projection measurements ${\{B}_k\}$. Since the systems possess quantum correlation, the quantum correlation will lead to another system disturbed when we measure a quantum system. Therefore, the two quantum expression of mutual information are no longer equal to each other. However the maximum of the second extension can be regarded as a measure of classical correlations $C({\rho}_{AB})$. It can be expressed as \cite{Ollivier, Henderson, S}:
\begin{eqnarray} C({\rho}_{AB})=S({\rho}_A)-\min_{{B}_k}S({\rho}_ {AB}|{\{B}_k\}),
\end{eqnarray}
where $S({\rho}_ {AB}|{\{B}_k\})=\sum_kP_kS({\rho}_ {AB}^k)$ is the conditional entropy of subsystem A \cite{S}, ${\rho}_ {AB}^k=({I}\otimes{B_k}){\rho}_ {AB}({I}\otimes{B_k})/P_k$ and  $P_k =Tr({I}\otimes{B_k}){\rho}_ {AB}({I}\otimes{B_k})$. The minimum value in eq. (2) arises from the complete set of projection measurements ${\{B}_k\}$. The minimum difference between the two quantum versions of mutual information is equal to QD namely ${{D}(\rho_{AB})}$. It can be written as \cite{Ollivier, Henderson}:
\begin{eqnarray} D({\rho}_ {AB})=I({\rho}_ {AB})-C({\rho}_ {AB}),
\end{eqnarray}
In order to study the two-qubit entanglement dynamics, we use Wooter's concurrence \cite{Wootters}. For two qubits, the concurrence is calculated from the density
matrix $\rho$ for qubits A and B:
\begin{eqnarray} C({\rho})=max{\{0,\sqrt{\lambda_1}-\sqrt{\lambda_2}-\sqrt{\lambda_3}-\sqrt{\lambda_4}}\},
\end{eqnarray}
Where the quantities  ${\lambda_i}$ are the eigenvalues in decreasing order of the matrix ${\xi}$:
\begin{eqnarray} \xi=\rho(\sigma_y\otimes\sigma_y)\rho^*(\sigma_y\otimes\sigma_y),
\end{eqnarray}
Where $\rho^*$ denotes the complex conjugation of $\rho$ in the standard basis $|00\rangle,|01\rangle,|10\rangle,|11\rangle$ and $\sigma_y$ the Pauli matrix.

\section{Schatten 1-norm GQD}
 We consider a bipartite system AB in a Hilbert space  ${H={H_A}\otimes{H_B}}$. The system is determined by quantum states characterized by density operators ${\rho\in{B(H)}}$, where B(H) is the set of bound, positive-semidefinite operators acting on H with ${Tr[\rho]=1}$. The 1-norm GQD between A and B is defined through the trace distance between ${\rho}$ and the closest classical- quantum state ${\rho_c}$ \cite{ Paula, B. Aaronson, Aaronson, T. Nakano},  reading
\begin{eqnarray} D_G({\rho})=\min_{\Omega_{0}}{{\|\rho-\rho_c\|}_1},
\end{eqnarray}
where ${\|X\|_1=Tr{[\sqrt{X^\dag X}]}}$ is the 1-norm (trace norm) and ${\Omega_0}$   is the set of classical-quantum states.

 In the certain case of two- qubit Bell diagonal states, whose density operator possess the form
 \begin{eqnarray} \rho={\frac{1}{4}}{[{I}\otimes{I}+\vec{c}.({\vec{\sigma}}\otimes{\vec{\sigma}})]},
\end{eqnarray}
Where I is the identity matrix, ${\vec{c}=(c_1,c_2,c_3)}$ is a three- dimensional vector such that ${-1\leq{c_i}\leq1}$ and ${\vec{\sigma}=(\sigma_1,\sigma_2,\sigma_3)}$ is a vector composed by Pauli matrices.

, 1- norm GQD can be analytically computed, yielding \cite{Paula}
\begin{eqnarray}D_G({\rho})=int[|c_1|,|c_2|,|c_3|],
\end{eqnarray}
where ${int[|c_1|,|c_2|,|c_3|]}$ is the intermediate result among the absolute values of the correlation functions ${c_1,c_2}$ and ${c_3}$.
\subsection{one- Norm GQD Under Decoherence}
 We will consider the evolution of a quantum state ${\rho}$ as described by a trace- preserving quantum operation ${\varepsilon(\rho)}$ which is given by
\begin{eqnarray}\varepsilon({\rho})=\Sigma_{i,j}({{E_i}\otimes{E_j}}){\rho}({{E_i}\otimes{{E_j}^\dag}}),
\end{eqnarray}
where ${\{E_k\}}$ is the set of Kraus operators corresponding with a decohering process of a single qubit, with the trace- preserving condition namely ${\Sigma_k{{E_k}^\dag{E_k}=I}}$.
In the case of bit flip noise one has following Kraus operators:
\begin{eqnarray}E_0={\sqrt{1-\frac{p}{2}}I}, E_1={\sqrt{\frac{p}{2}}\sigma_1} ,
\end{eqnarray}
Similarly, the Kraus operators for generalized amplitude damping noise are given by
\begin{eqnarray}E_0={\sqrt{p}}\left(
\begin{array}{cccc}
1 && 0 \\
0 && {\sqrt{1-\gamma}}\\

\end{array}
\right), E_1={\sqrt{p}}\left(
\begin{array}{cccc}
0 && {\sqrt{\gamma}}\\\nonumber
0 && 0\\

\end{array}
\right) \\
E_2={\sqrt{1-p}}\left(
\begin{array}{cccc}
{\sqrt{1-\gamma}} && 0 \\ \nonumber
0 && 1\\
\end{array}
\right), E_3={\sqrt{1-p}}\left(
\begin{array}{cccc}
0 &&  0\\\nonumber
{\sqrt{\gamma}} && 0\\

\end{array}
\right) \\
\end{eqnarray}
Where p and ${\gamma}$ refers to decoherence probability. The BF channel preserve the Bell- diagonal form of the density operator ${\rho}$. However for the GAD channel, the Bell- diagonal form is preserve for every ${\gamma}$ and ${p=\frac{1}{2}}$. In this situation, the values of correlation vector are given in Table I.

\begin{table}[t]
\caption{\label{blobs}Correlation functions for the quantum operation: bit flip (BF) and generalized amplitude damping (GAD). For GAD, we fixed p=1/2. Where, $c'$s mean the evolved coefficients of the Bell-diagonal state}
\begin{indented}
\item[]\begin{tabular}{@{}lllll}
\br
channel& ${\acute{c_1}}$&&${\acute{c_2}}$&${\acute{c_3}}$\\
\mr
\verb"BF"&${c_1}$        &&${c_2(1-p)^2}$&${c_3(1-p)^2}$\\
\verb"GAD"&${c_1(1-\gamma)}$        &&${c_2(1-\gamma)}$&${c_3(1-\gamma)^2}$\\

\br
\end{tabular}
\end{indented}
\end{table}

1- norm GQD can be obtained straightforwardly, since the Bell- diagonal form is preserved . It suffices instead of ${c_i}$ in equation (9), we put ${c_i^\prime}$ from Table I \cite{B. Aaronson, Aaronson, T. Nakano, Montealegre}.

Then, we can obtain the 1- norm GQD decay as a function of decoherence probabilities.
 \section{The Heisenberg XXX model}
Here we analyze the thermal Bell diagonal states (BDSs) in a two qubit XXX system. The Hamiltonian of the Heisenberg XXX model
with no external magnetic field and only nearest neighbor interactions is \cite{Rigolin}
\begin{eqnarray}H={\frac{J}{4}(\sigma_z^1\sigma_z^2+\sigma_y^1\sigma_y^2+\sigma_x^1\sigma_x^2)},
\end{eqnarray}
Where J is coupling constant, here ${J>0}$ correspond to the anti- ferromagnetic case; ${J<0}$ correspond to the ferromagnetic case; ${\sigma_x, \sigma_y, \sigma_z}$ are the pauli matrices, and ${\hbar=1}$. The four eigenvectors of the Hamiltonian are the four Bell states: ${H|\varphi^\pm\rangle=\lambda_{\varphi^\pm}|{\varphi^\pm}\rangle}$ and ${H|\psi^\pm\rangle=\lambda_{\psi^\pm}|{\psi^\pm}\rangle}$, where ${|\varphi^\pm\rangle=(\frac{1}{\sqrt2})(|00\rangle\pm|11\rangle)}$, ${|\psi^\pm\rangle=(\frac{1}{\sqrt2})(|01\rangle\pm|10\rangle)}$. And the eigenvalues of the Hamiltonian are: ${\lambda_{\varphi^{\pm}}=\frac{J}{4}, \lambda_{\psi^{+}}=\frac{J}{4}}$ and ${\lambda_{\psi^{-}}=\frac{-3J}{4}}$.
The density matrix which describes a system in thermal equilibrium at temperature T is ${\rho={\exp(-{H}/{kT})}/Z}$, where ${Z=Tr[{\exp(-{H}/{kT})}]}$ is the partition function and k is Boltzmann's constant. Hamiltonian (12) gives the following thermal state written in the standard basis,${\{|00\rangle, |01\rangle, |10\rangle, |11\rangle\}}$.
\begin{eqnarray}
\rho={\frac{1}{Z}} \left(
\begin{array}{cccccccccccccccc}
\rho_{11} && 0 && 0 && 0\\
0 && \rho_{22} && \rho_{23} && 0\\
0 && \rho_{32} && \rho_{33} && 0\\
0 && 0 && 0 && \rho_{44}\\
\end{array}
\right).
\end{eqnarray}
The  matrix elements ${\rho_{11},  \rho_{22}, \rho_{23}, \rho_{32}, \rho_{33}, \rho_{44}}$ of ${\rho}$ are
\begin{eqnarray}\rho_{11}= \rho_{44}&=&\exp(-\alpha),  \rho_{22}= \rho_{33}={\exp(\alpha)}{\cosh(2\alpha)}\nonumber\\
 and    &\rho_{23}&= \rho_{32}=-{\exp(\alpha)}{\sinh(2\alpha)} \end{eqnarray}
Where the partition function ${Z=2(\exp(-\alpha)+\exp(\alpha)\cosh(2\alpha))}$.
It can be further expressed in the BDS form with the coefficients
\begin{eqnarray}
c_1=-{\frac{2}{Z}}(\exp(\alpha)\sinh(2\alpha)), c_2=c_1 ,    c_3=({\frac{4}{Z}}\exp(-\alpha)-1 ) , (\alpha={\frac{J}{4kT}})
\end{eqnarray}
 The quantum discord for the BDS (13) can be exactly calculated as \cite{S}
 \begin{eqnarray}
QD(\rho)=&\frac{1}{4}&[(1-c_1-c_2-c_3)\log_2{(1-c_1-c_2-c_3)}\nonumber\\
&+&(1-c_1+c_2+c_3)\log_2{(1-c_1+c_2+c_3)}\nonumber\\
&+&(1+c_1-c_2+c_3)\log_2{(1+c_1-c_2+c_3)}\nonumber\\
&+&(1+c_1+c_2-c_3)\log_2{(1+c_1+c_2-c_3)}]\nonumber\\
&-&\frac{1-c}{2}\log_2{(1-c)}-\frac{1+c}{2}\log_2{(1+c)}
\end{eqnarray}
with $c=\max\{|c_1|, |c_2|, |c_3|\}$

\section{Quantum correlation without decoherence}
The quantum discord, 1-norm GQD and entanglement are known to vary with the parameters J and T.
Firstly, let us start with the QD, for finite temperature ${T>0}$. QD can be easily computed from equation (16). Therefore for simplicity we ignore writing it here. Figure 1 shows QD variation with respect to J and T. We can see that, by growing the absolute value of coupling strength (J), QD increase and behaves conversely to the
increase of temperature T. In addition, from figure 1(b) we can see that QD decreases by increasing of temperature T.  By knowing ${c_i}$s  we can compute 1- norm GQD easily. Compared with variation of QD mentioned in Figure 1, the variation of 1-norm GQD has a similar behavior as shown in Figure 2 (in two cases quantum correlation in anti- ferromagnetic region is more than ferromagnetic region).

\begin{figure}
\includegraphics[width=3.5in]{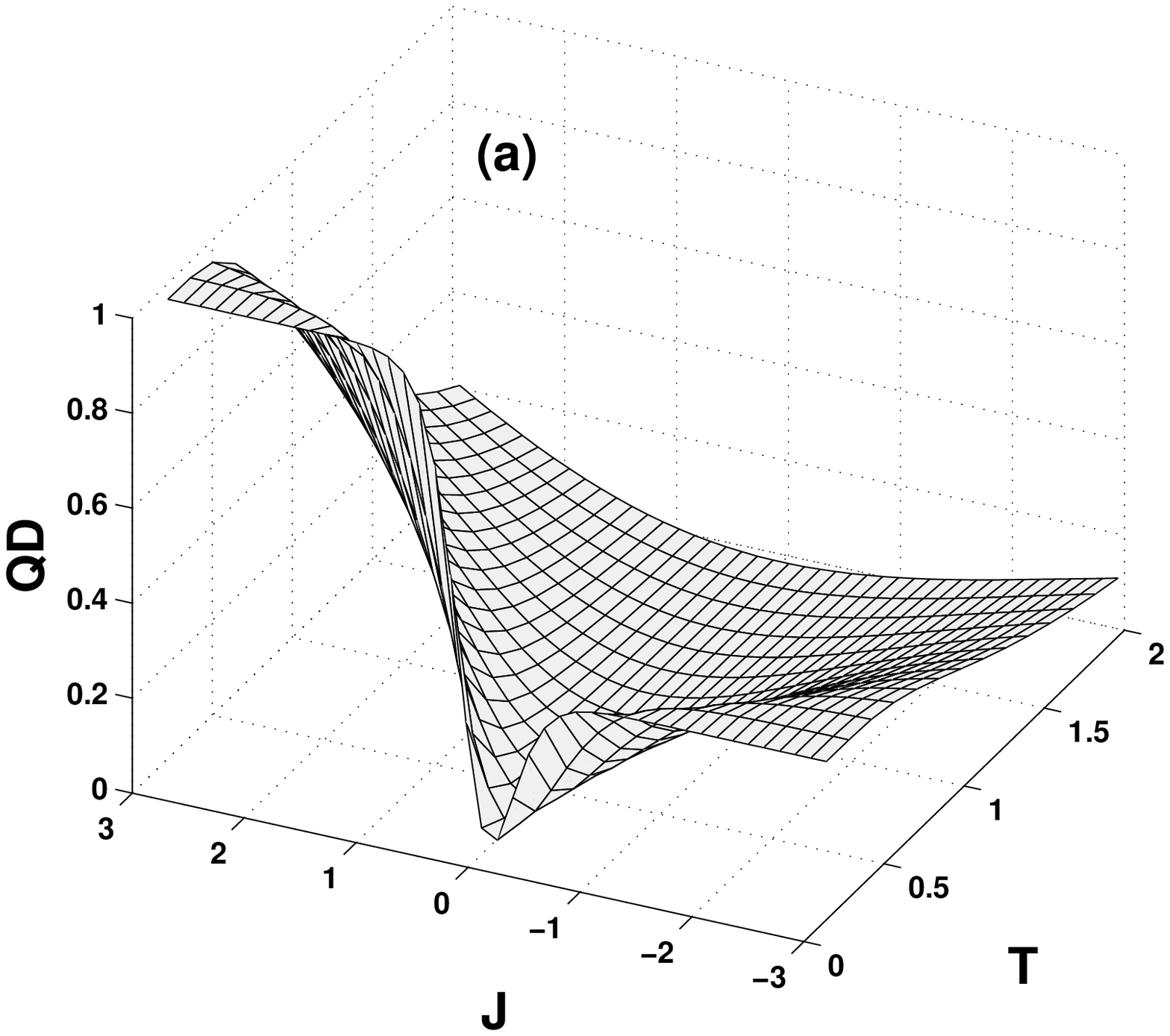}
\includegraphics[width=3in]{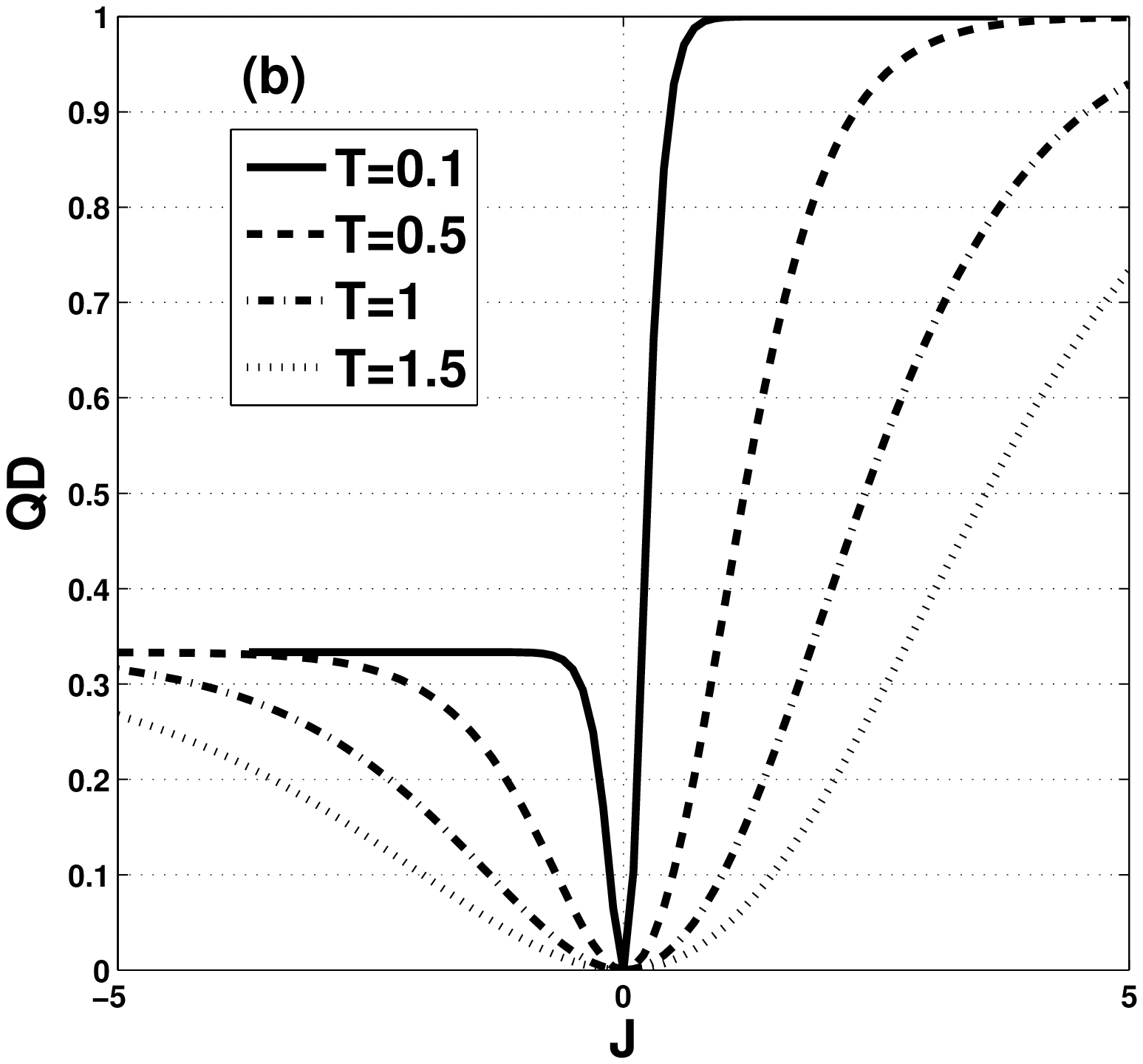}

\caption{(a) QD as a function of T and J; (b) QD as a function of J for different T.}
 \label{fig1}
\end{figure}

\begin{figure}
\includegraphics[width=3.5in]{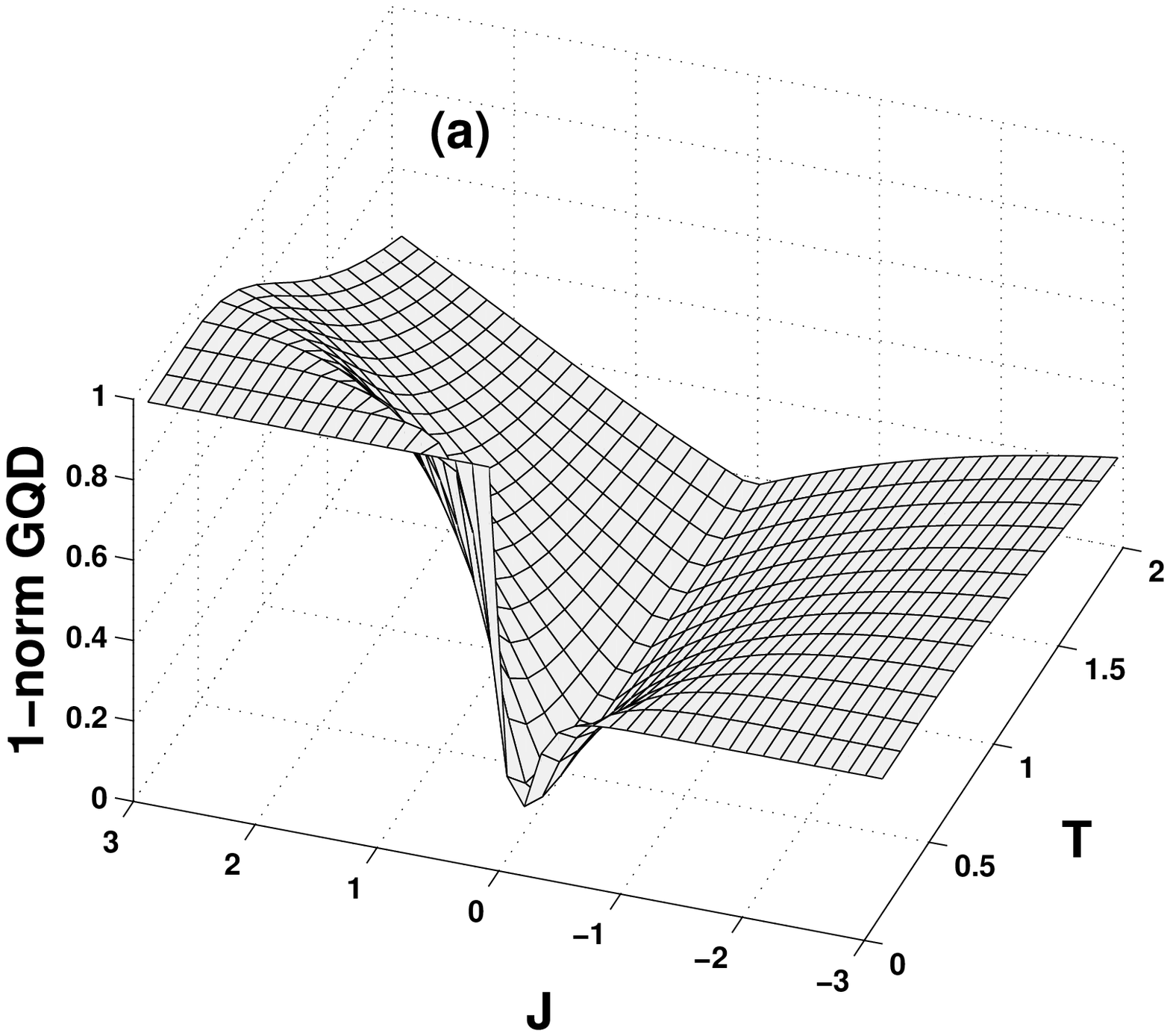}
\includegraphics[width=3in]{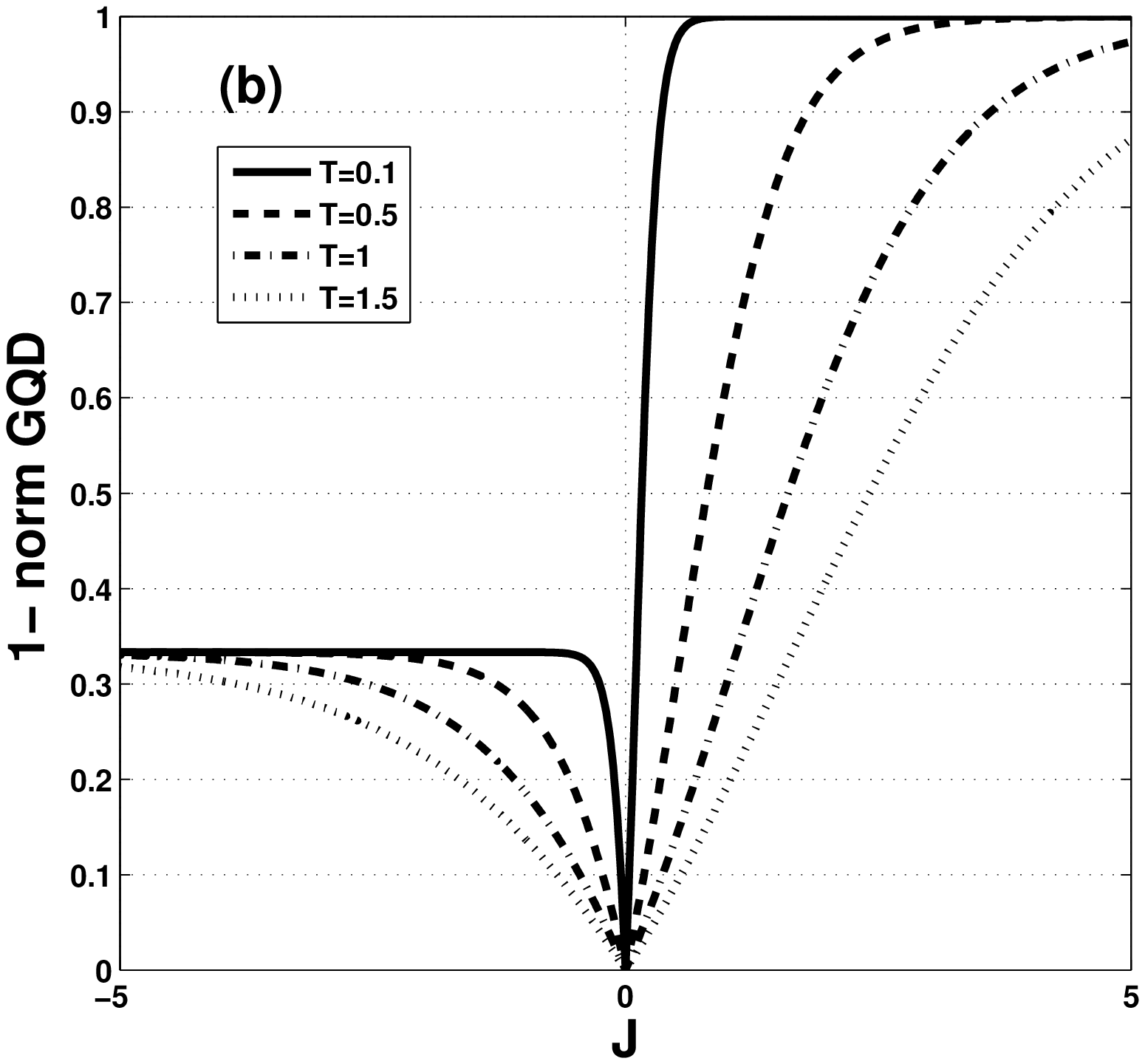}

\caption{(a) 1- norm GQD as a function of T and J; (b) 1- norm GQD as a function of J for different T.}
 \label{fig2}
\end{figure}
  In the case which decoherence is absent, concurrence for XXX Heisenberg spin chain is given by  \cite{Rigolin}\\
  \begin{eqnarray} C=\max(0, \frac{e^\alpha\sinh(2\alpha)-e^{-\alpha}}{e^\alpha\cosh(2\alpha)+e^{-\alpha}}),
\end{eqnarray}

  In figure 3(a) we have plotted concurrence as a function of T and different J. It can be observed that the higher the temperature is, the less the entanglement between two spins will be, which is consistent with QD and 1- norm GQD.
 When the temperature reaches some point (critical temperature $T_c$), the entanglement will disappear. It is notable that, for lesser J
 the region of T in which zero concurrence  appear is more. Figure 3(b) shows concurrence as a function of J and for different T. We can see that in contrary with
 QD and 1- norm GQD, concurrence in ferromagnetic region is always zero. It shows that the QD and 1- norm GQD are more general quantum correlations than entanglement.
The thermal concurrence has been studied by G. Rigolin \cite{Rigolin}, which shows that concurrence is zero for the ferromagnetic XXX model.

In Figure 4 the behaviors of QD, 1-norm GQD and concurrence at certain temperature are analyzed.   QD, 1-norm GQD and concurrence are plotted as a function of J at certain T (T=1). It can be seen that, 1-norm GQD is always higher than QD, which is consistent with that given by F.M. Paula et al. \cite{Paula}. Also, anti- ferromagnetic coupling $(J>0)$ can endure more quantum correlation. Although for $J<0$, there is no entanglement , QD and 1- norm GQD exist.
\begin{figure}
\includegraphics[width=3in]{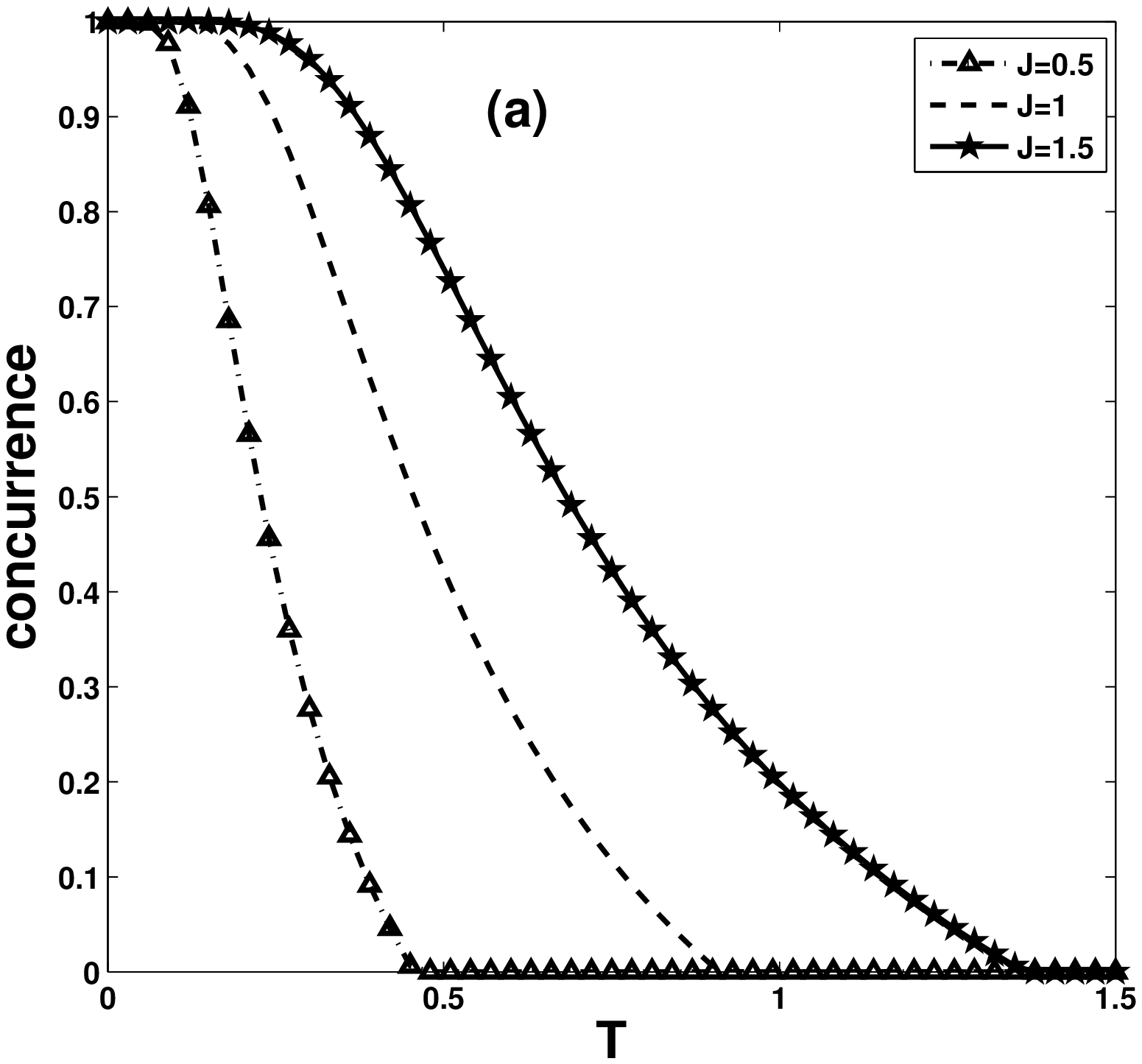}
\includegraphics[width=3in]{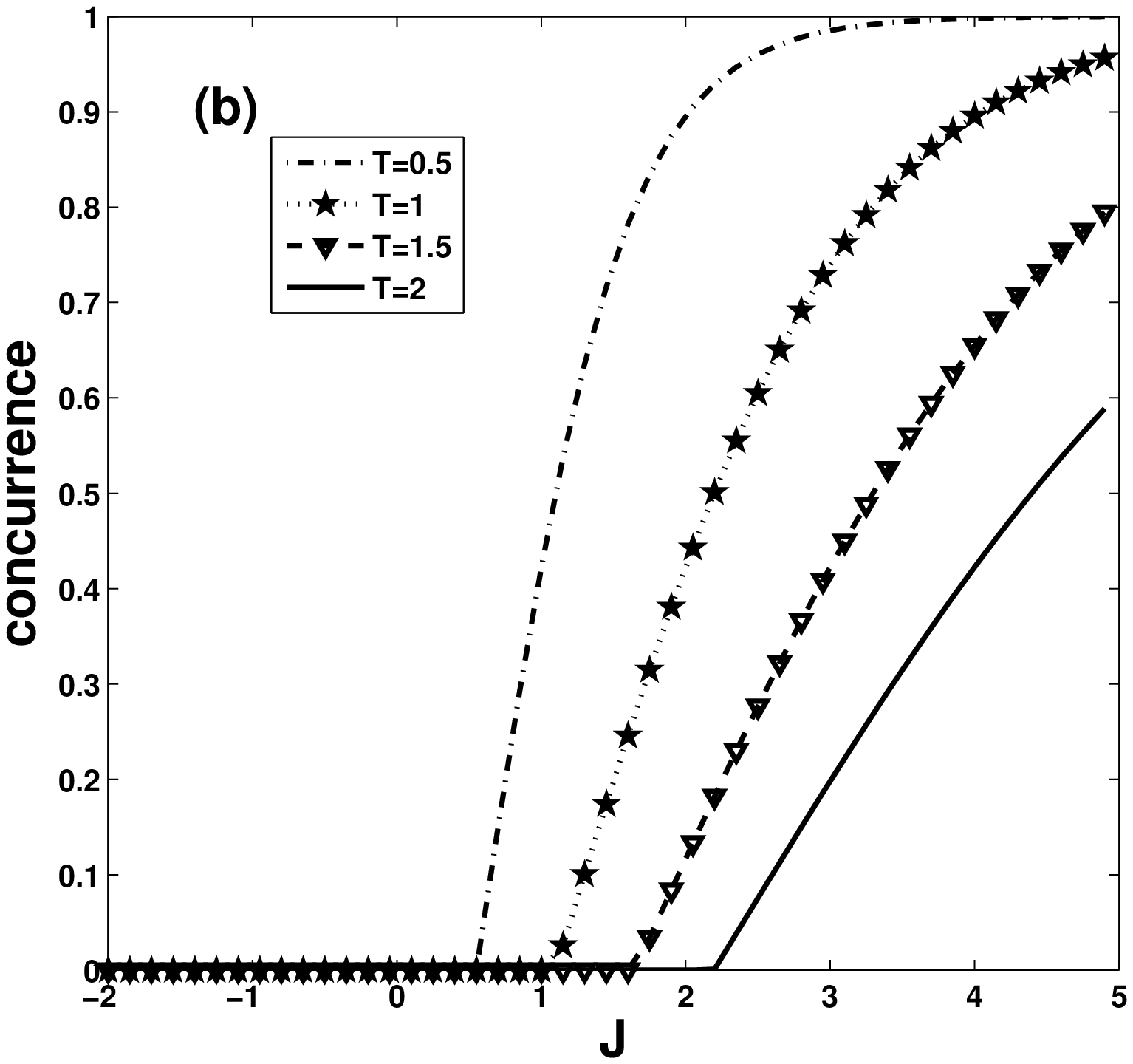}
\caption{(a) concurrence as a function of T for different J; (b) concurrence as a function of J for different T.}
 \label{fig3}
\end{figure}
\begin{figure}
\includegraphics[width=3in]{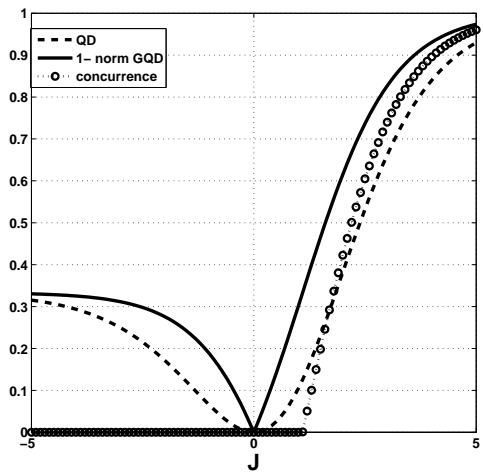}
  \caption{comparison of QD, 1- norm GQD and concurrence as a function of J at a certain temperature (T=1).}\label{fig4}
\end{figure}

\section{Quantum correlation under decoherence}
\subsection{Bit flip channel case}

Now we consider QD, 1-norm GQD and concurrence of XXX model in the presence of various  noises. First of all, we consider the effect of BF channel on the XXX Hiesenberg chain.
Again like pervious section, we ignore from writing expression of QD and 1- norm GQD (which can be easily computed with inserting ${c_i^\prime}$ from Table I in equation 14). However concurrence for this case is as follow
   \begin{eqnarray} C(\rho)=2\max\{0, |\frac{2}{Z}e^\alpha\sinh(2\alpha)(1+(1+p)^2)|\\\nonumber
   -(1+(\frac{4}{Z}e^{-\alpha}-1)(1-p)^2\},
\end{eqnarray}

Here, we show the dynamics of QD, 1-norm GQD and concurrence qualitatively with respect to different parameters under the effect of  BF channel as shown in the following figures.
Figures 5(a), 5(b) and 5(c) show the behavior of the two- qubit QD, 1-norm GQD and concurrence under BF noise. As we can see from Figure 5, entanglement sudden death occurs while QD and 1- norm GQD vanishes only in asymptotic limit. It is remarkable that, 1-norm GQD is always existent while QD and concurrence vanishes in ferromagnetic regions. At high temperature the entanglement will disappear. We can compare QD, 1- norm GQD and concurrence as a function of coupling constant in fix T (which we consider T=1) under effect of BF noise (see figure 6). As we can see form figure 6, the overall shape is not like the case in which decoherence was absent. Specially, 1- norm GQD is not higher than QD and these correlations have a different order as a function of J. Such a behavior is in contrast with the previous results which provided by F. M. Paula et al. \cite{Paula}. Therefore, we claim that there is no simple relative ordering between QD and 1- norm GQD.\\

\begin{figure}
  \includegraphics[width=3in]{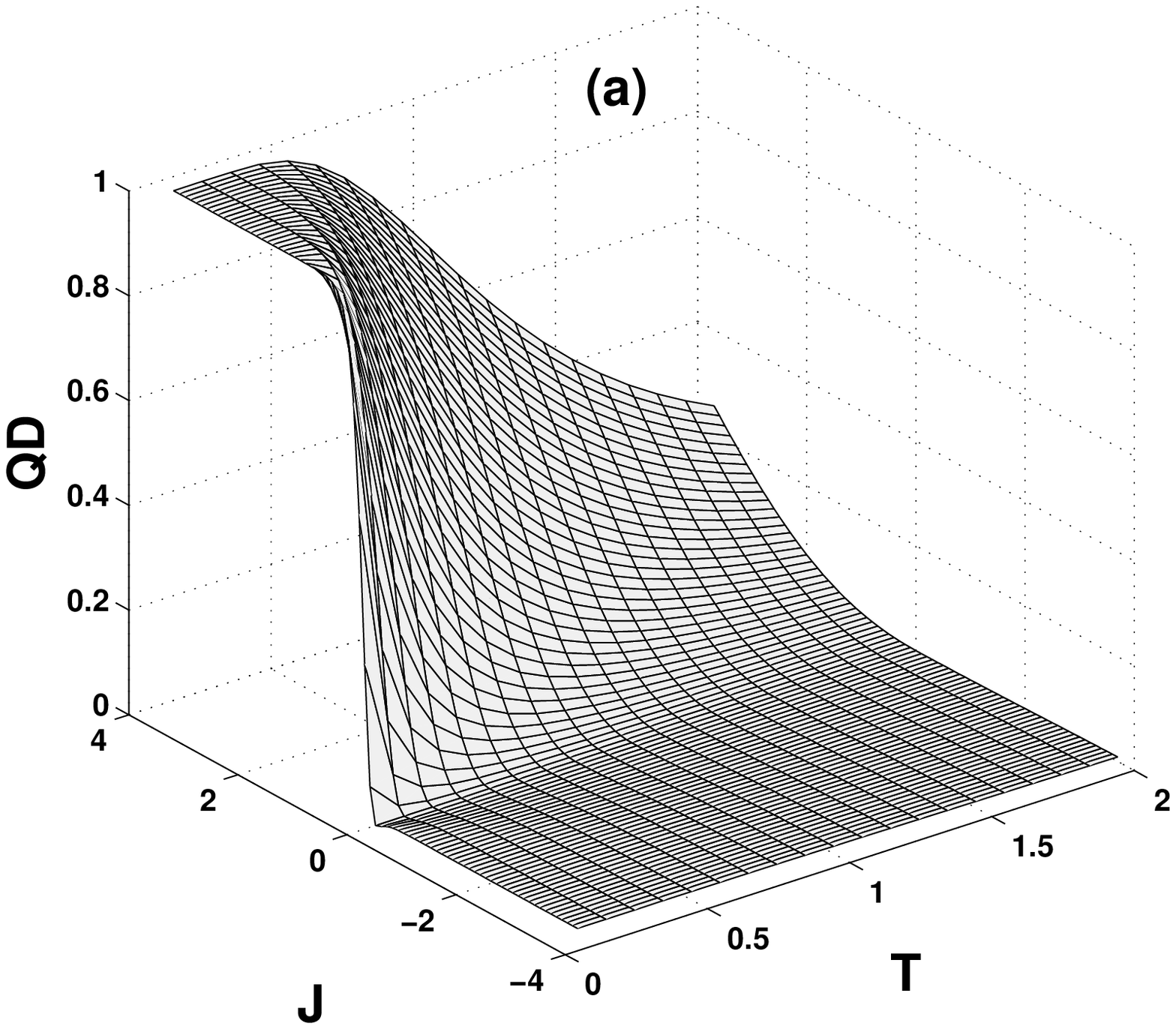}\\
\includegraphics[width=3in]{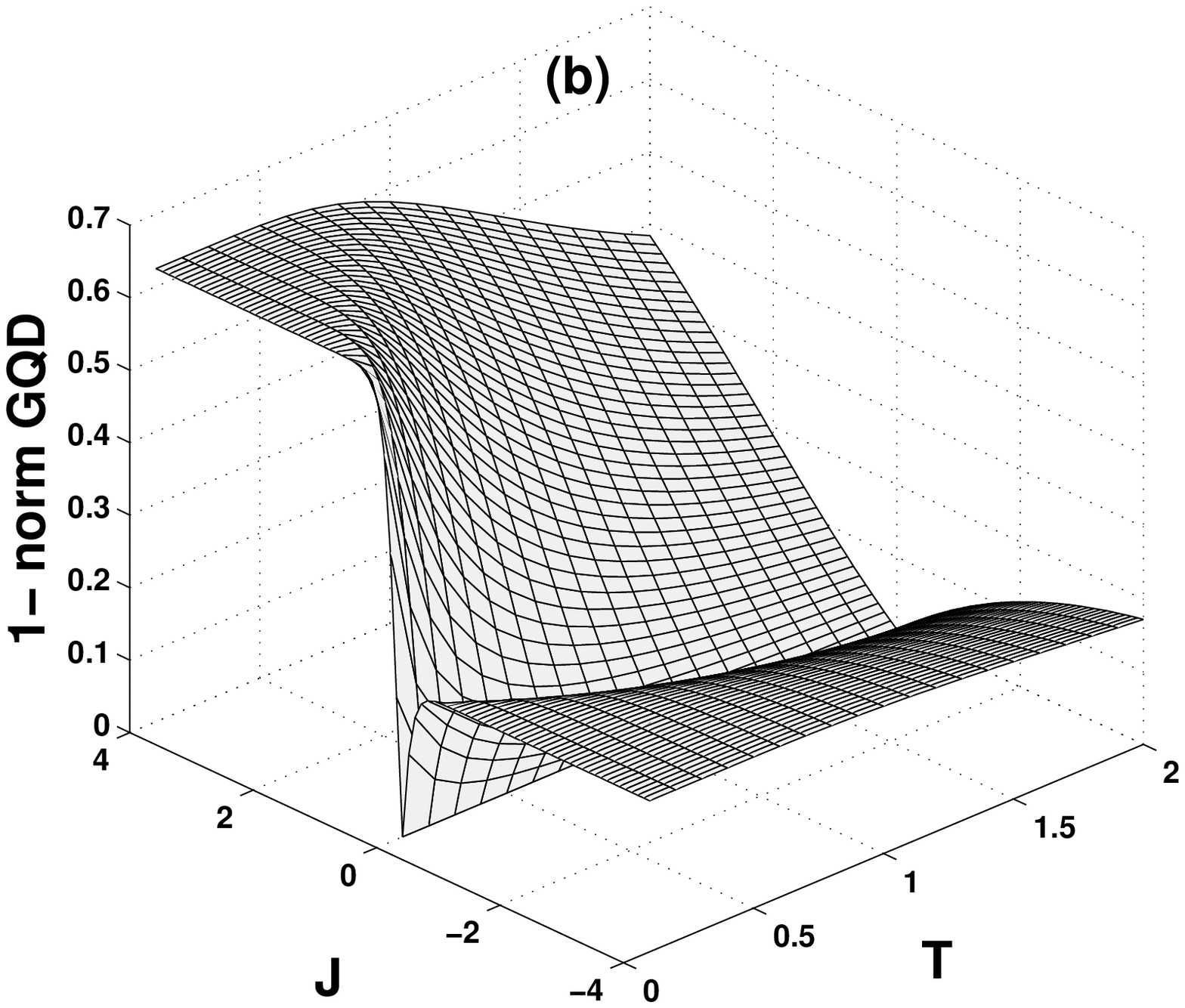}
\includegraphics[width=3in]{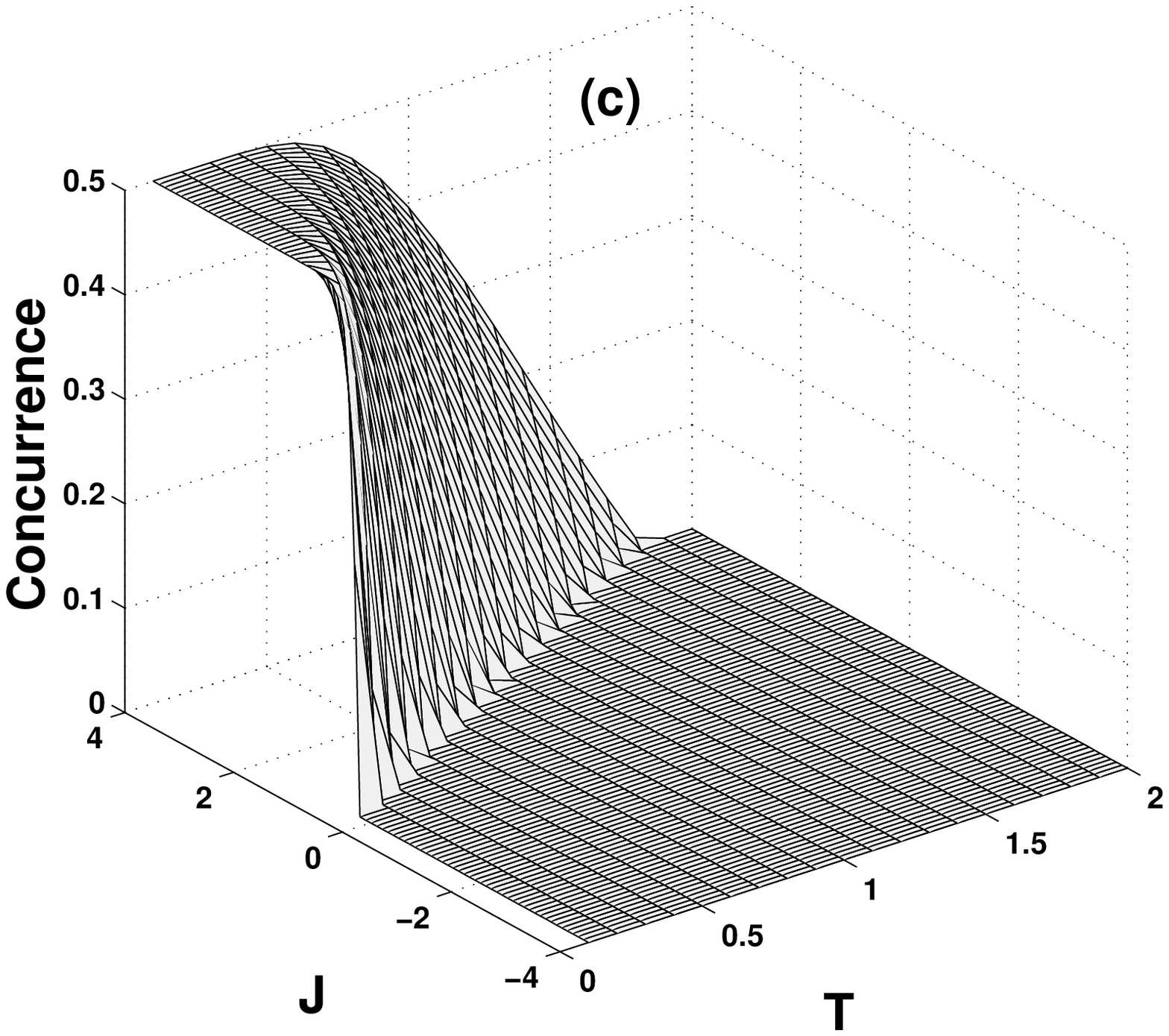}
\caption{Quantum correlations as a function of J and temperature T under the effect of BF nchannel when p=1/2 (a) QD, (b) 1- norm GQD and (c) concurrence respectively.}
 \label{fig5}
\end{figure}

\begin{figure}
  \includegraphics[width=3in]{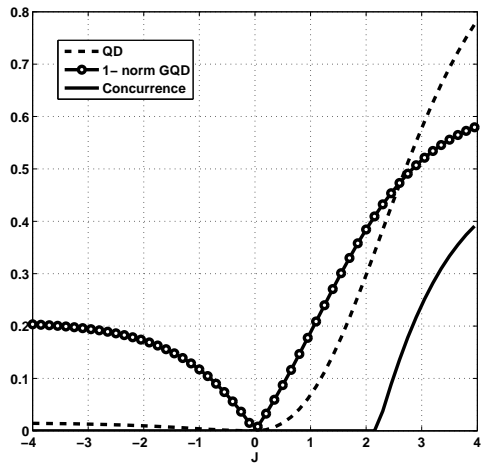}\\
  \caption{comparison of QD, 1- norm GQD and concurrence as a function of J in a certain T (T=1) under BF channel when p=1/2.}\label{fig.6}
\end{figure}

\subsection{Generalized amplitude damping channel case}

For XXX model in the presence of generalized amplitude damping (GAD) channel, concurrence can be written as follow
  \begin{eqnarray} C(\rho)=2\max\{0, |\frac{4}{Z}e^\alpha\sinh(2\alpha)(1-\gamma)|-(1+(\frac{4}{Z}e^{-\alpha}-1)(1-\gamma)^2\},
\end{eqnarray}

   The QD, 1-norm GQD and concurrence for the thermal state (16) as function of temperature and J are plotted in Figure 7. One can see that the overall behaviors  of discord and 1-norm GQD are similar.  When absolute value of J increase, both of them grow rapidly. It is hold for concurrence, however  the distinct difference between concurrence and discord is: QD and 1-norm GQD vanishes in an asymptotic way, but concurrence behaves as sudden death at a finite critical temperature $T_c$. The lifetime of QD and 1- norm GQD is significantly longer than that of entanglement as measured by concurrence. QD and 1- norm GQD may be more robust against decoherence. In figure 8 we can compare the behavior of QD, 1- norm GQD and concurrence as a function of J (with T=1 and ${\gamma=1/2}$). we can see that, in contrary with the BF channel 1- norm GQD is larger than QD in the case of GAD channel. However, it has similar behavior with QD, which confirms that 1- norm is a good measure for quantum correlation.

\begin{figure}
\includegraphics[width=3in]{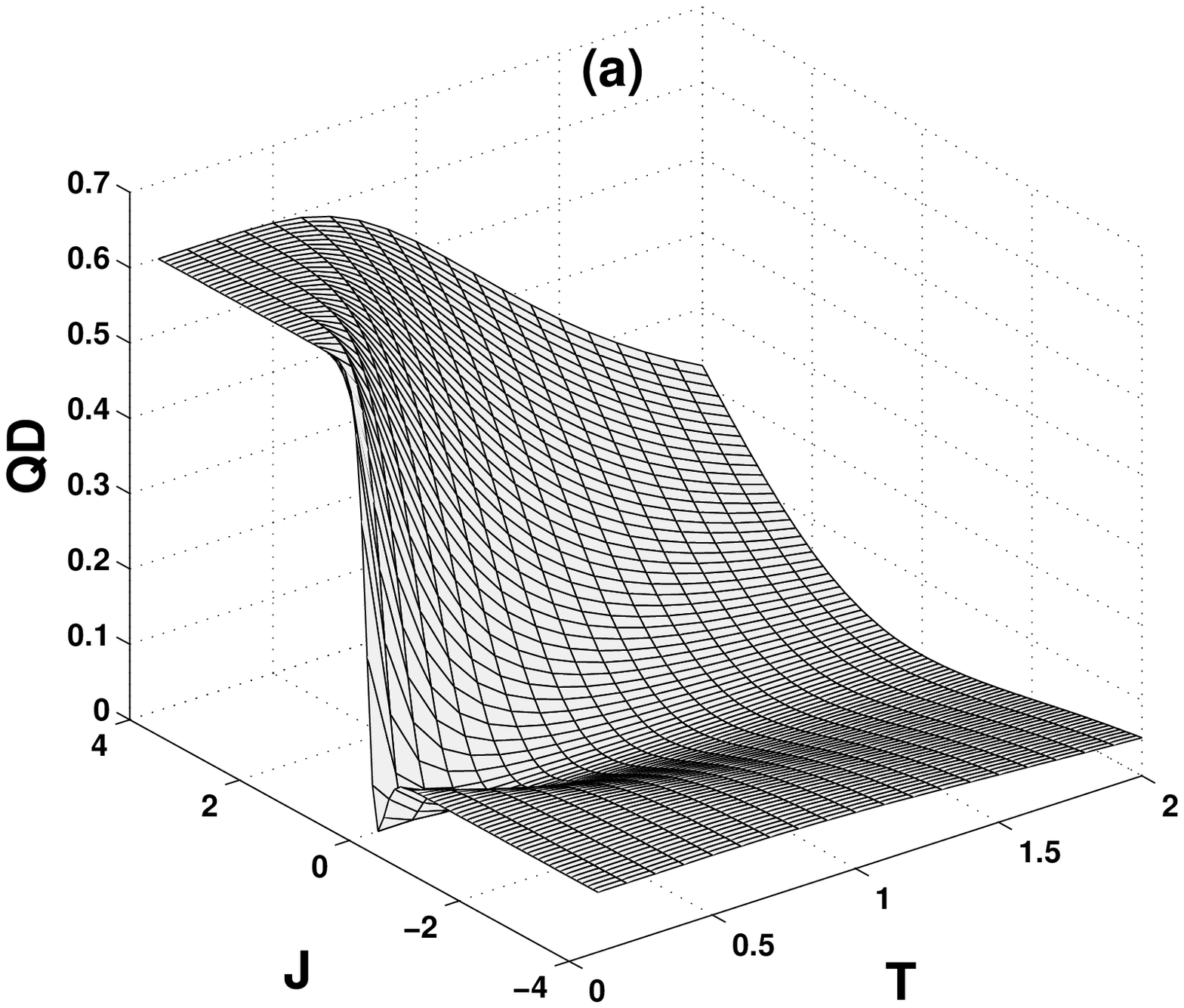}
\includegraphics[width=3in]{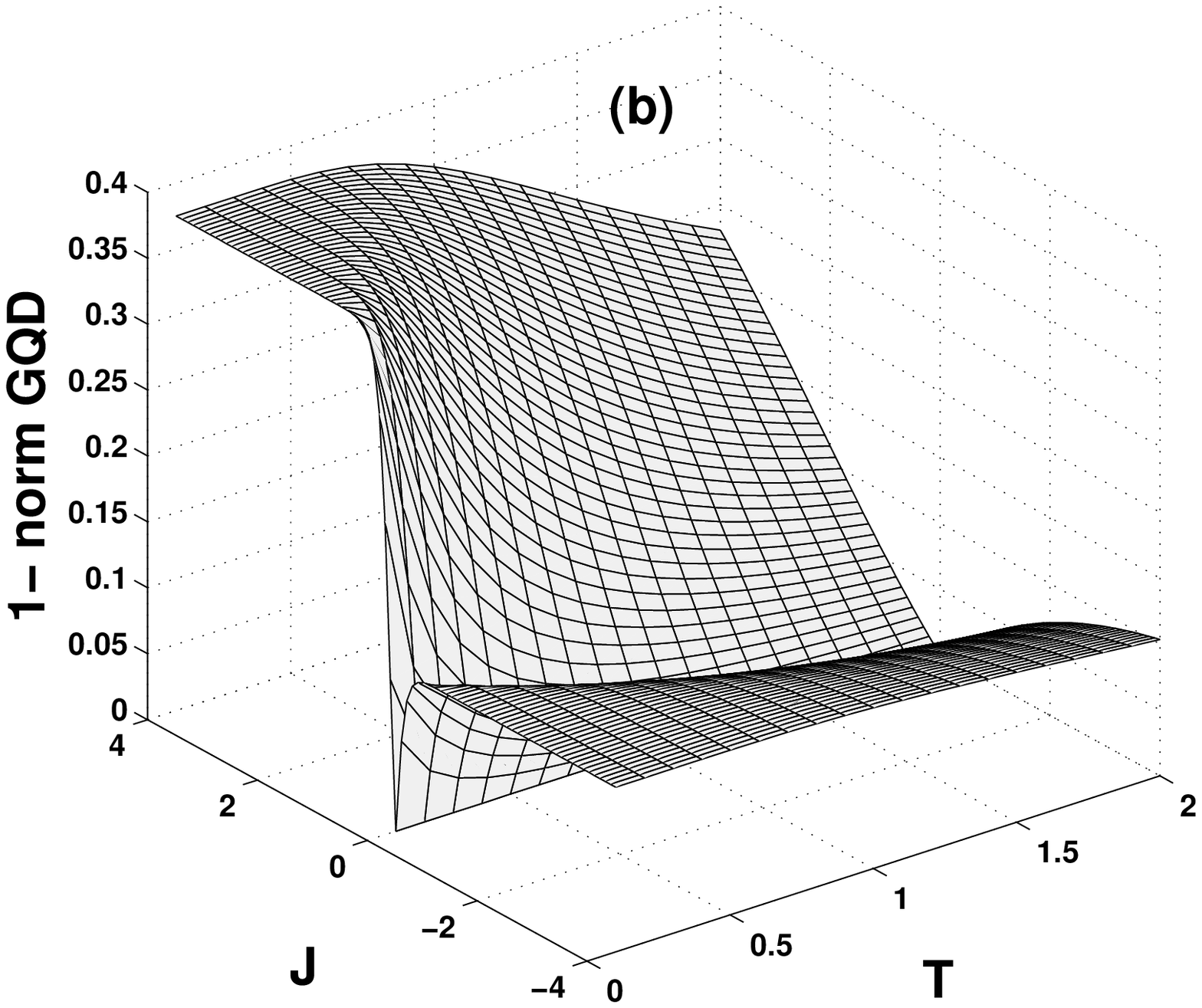}
\includegraphics[width=3in]{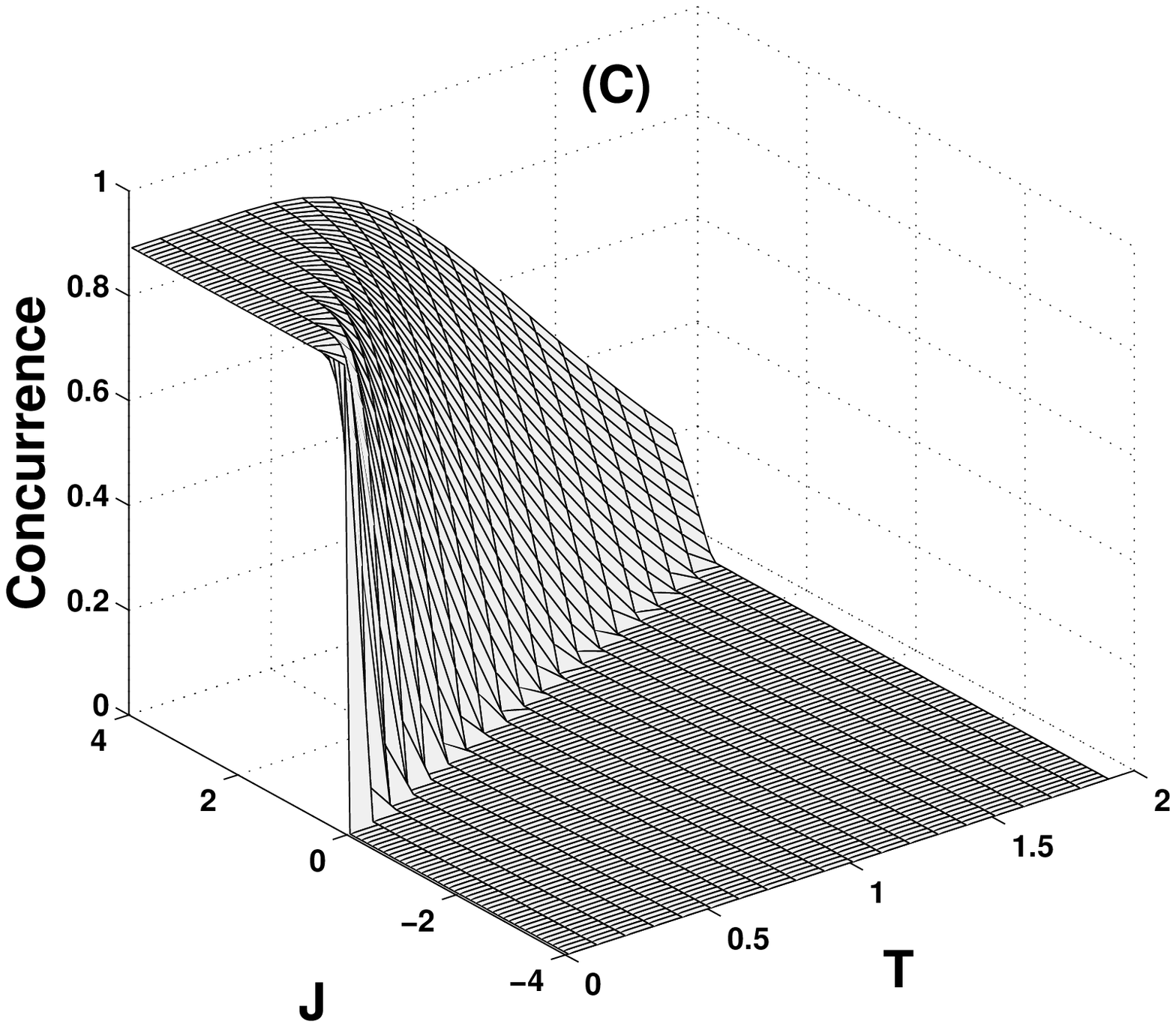}
\caption{Quantum correlations as a function of J and temperature T under the effect of GAD channel when ${\gamma=1/2}$ (a) QD, (b) 1- norm GQD and (c) concurrence respectively.}
 \label{fig7}
\end{figure}

\begin{figure}
  \includegraphics[width=3in]{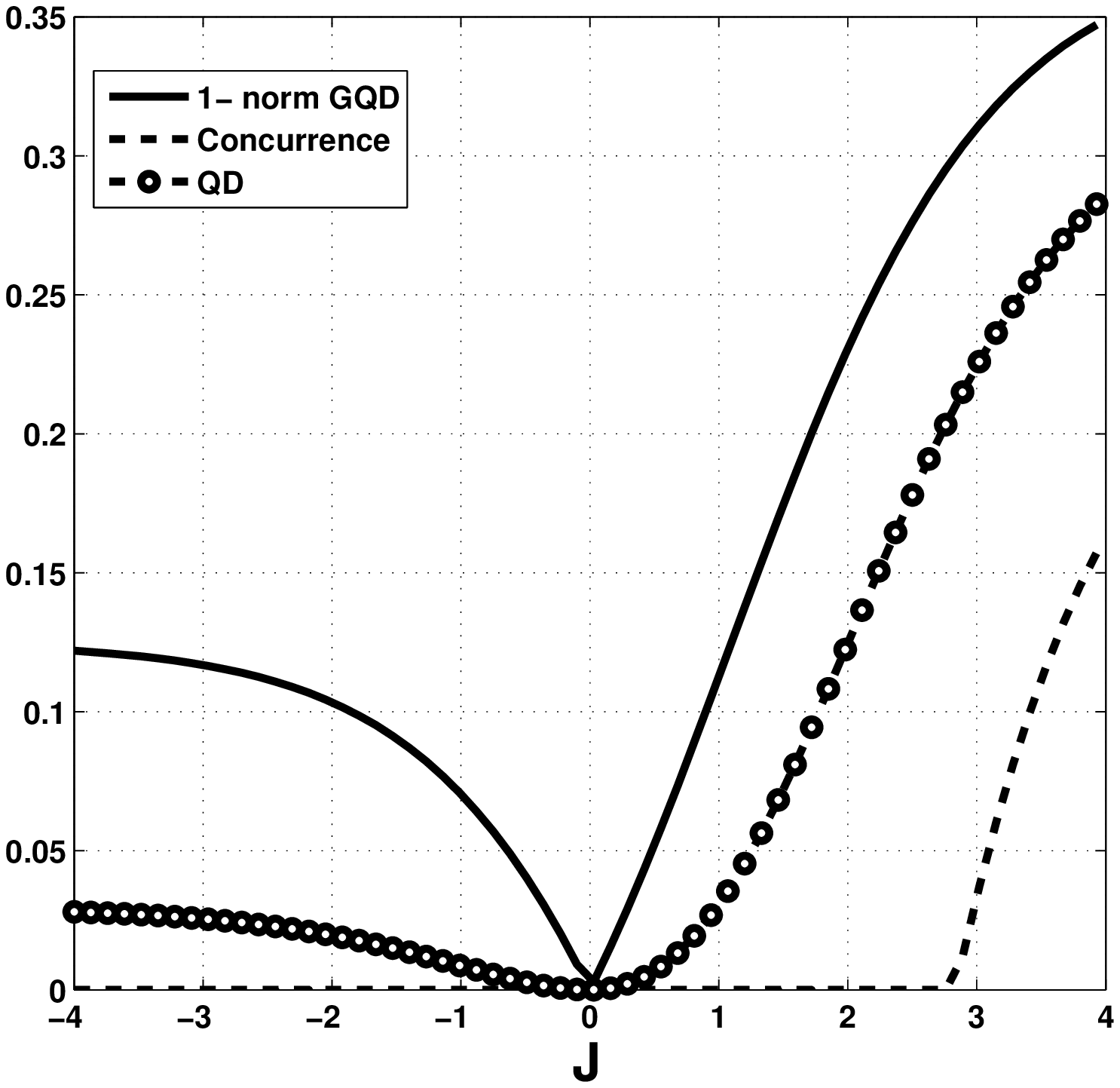}\\
  \caption{comparison of QD, 1- norm GQD and concurrence as a function of J in a certain T (T=1) when ${\gamma=1/2}$.}\label{fig8}
\end{figure}

\section{Conclusion}
 In this paper, we have studied the variation of QD, 1- norm GQD and concurrence in Heisenberg XXX spin system  with  no external magnetic field and only nearest neighbor interactions. The dependencies of these three quantities on the temperature and coupling constant in the absent and present of noises are given in detail. More important, we have compared these quantities and found no definite link between them, so that 1- norm GQD may be smaller or larger than QD.
  QD and 1- norm GQD are always existent while the entanglement disappears in some region, which indicate the importance of QD and 1- norm GQD. Specially in ferromagnetic region entanglement is always zero even in the absence of noises. Moreover, entanglement decreases more rapidly than QD and 1-norm GQD and entanglement sudden death occurs while QD and 1- norm GQD vanish only in the asymptotic limit. For an XXX model without magnetic field, anti- ferromagnetic coupling
($J>0$) can endure more quantum correlation.  Although for $J<0$, there is no entanglement, QD and 1- norm GQD exist. Therefore, anti- ferromagnetic coupling can be more helpful than ferromagnetic coupling for quantum correlations.

\end{document}